\documentclass[10pt,pra,twocolumn,notitlepage,floatfix,superscriptaddress,nofootinbib]{revtex4-1}
\usepackage{graphicx}
\usepackage{bbm}
\usepackage{epstopdf,float}
\usepackage[colorlinks]{hyperref}
\hypersetup{
 colorlinks=true,
 citecolor=magenta,
 linkcolor=cyan,
 urlcolor=black}
\usepackage{amssymb}
\usepackage{times,bm}
\usepackage{amsmath}
\usepackage{mathrsfs}
\usepackage{siunitx}
\makeatletter
\def\Let@{\def\\{\notag\math@cr}}
\makeatother
\usepackage{amsthm,amsfonts}
\usepackage[caption=false]{subfig}
\newcommand{\bra}[1]{\langle #1|}
\newcommand{\ket}[1]{|#1\rangle}

\renewcommand{\sec}[1]{\hyperref[sec:#1]{Sec.~\ref{sec:#1}}}

\newcommand{\fig}[1]{\hyperref[fig:#1]{Fig.~\ref{fig:#1}}}
\newcommand{\tab}[1]{\hyperref[fig:#1]{Table~\ref{tab:#1}}}

\newcommand{\Tr}[1]{\textrm{Tr}\left\{#1 \right\}}

\allowdisplaybreaks

\begin{document}
\title{Universal quantum computing with thermal state bosonic systems}
\author{Kevin Marshall}
\author{Daniel F. V. James}
\affiliation{Center for Quantum Information and Quantum Control (CQIQC),
 Department of Physics, University of Toronto, 60 Saint George Street, Toronto, Ontario, M5S 1A7, Canada}
\author{Alexandru Paler}
\affiliation{Linz Institute of Technology, Johannes Kepler University, Linz, 4040, Austria}
\author{Hoi-Kwan Lau}\email[Email address:]{ hklau.physics@gmail.com}
\affiliation{Institute for Molecular Engineering, University of Chicago, 5640 South Ellis Avenue, Chicago, Illinois 60637, U.S.A.}
\begin{abstract}
Recent development of mixed-state encoding (MSE) allows pure-state logical information to be encoded by a bosonic (continuous-variable) system in \textit{mixed} physical state.  Despite interest due to its counter-intuitiveness, the utility of the current MSE scheme is limited due to several operational drawbacks, namely redundant information carrier, probabilistic initialisation, and requirement of discrete-variable measurement.  
In this work, we present a simplified MSE that does not suffer from any of these drawbacks.
Specifically, our protocol encodes each qubit by only one mixed-state bosonic mode, and the logical basis can be deterministically initialised from thermal equilibrium.  
All logical operations of this encoding can be performed with continuous-variable interaction and measurement only.  Without the necessity of ground state cooling, our proposal could broaden the set of candidate systems for implementing quantum computers, 
and reduce the reliance on demanding refrigerating facility for current quantum computing architectures.  
Additionally, our protocol can enhance the noise tolerance of logical qubit even if the system can be efficiently cooled.
\end{abstract}
\date{\today}
\maketitle

\section{Introduction}\label{sec:intro}

Quantum computers are widely believed to offer dramatic speedup in a variety of applications, such as solving algebraic problems, physical simulation, and machine learning \cite{book:NielsenChuang, childs10, georgescu14, 2017Natur.549..195B}.
In most quantum computing algorithms, the basic unit of information is a qubit, which is generally a superposition of two logical values.  
For a physical system to be a candidate of quantum computer, it should meet at least two conditions \cite{DiVincenzo:2000vw}: first, it exhibits well characterised (and decoherence-robust) degrees of freedom for representing qubits; second, physical controls should be available to implement all logical operations, including initialisation, logic gates, and information readout. 

In the last two decades, several promising quantum computer candidates have been recognised \cite{Ladd:2010kq}; they can be roughly divided into two categories: discrete-variable (DV) and continuous-variable (CV) systems.  In DV systems, each physical degree of freedom exhibits finite but individually addressable states.  Examples of DV systems include the internal states of trapped ions \cite{Haffner:2008tg} and the spin states in diamond colour centres \cite{Doherty:2013tv}.
In contrast, each physical degree of freedom of a CV system behaves as a bosonic quantum mode (qumode) \cite{braunstein05, weedbrook12}.  Each qumode exhibits effectively infinite eigenstates, but individually addressing a specific state could be challenging.  Examples of CV systems includes: cavity and travelling photonic modes \cite{2007RvMP...79..135K}, superconducting resonators \cite{2017NatCo...8...94H, 2017NatCo...8.1904N}, mechanical oscillators \cite{Leibfried:2003tg, OConnell:2010br, Poot:2012fh}, and spin ensembles \cite{Rabl:2006if, Wesenberg:2009es}.
When comparing to DV systems, CV systems are ubiquitous, and some offer the advantage of long coherence time \cite{2013ApPhL.102s2604R}, simple error correction \cite{mirrahimi14, binomial16}, and efficient generation of large scale multi-partite entanglement \cite{thousand_mode, million_mode}.

At the beginning of quantum computation, the physical system has to be initialised as a logical basis \cite{DiVincenzo:2000vw}.  
Unlike DV systems, in which the logical bases are usually two physical eigenstates, there is no natural choice of CV encoding states.
In the literature, numerous CV encodings have been proposed to represent logical bases as, e.g. Fock states, coherent states, cat states, etc \cite{chuang95,chuang96,knill01,ralph03,lund08,An09,leghtas13,mirrahimi14,gottesman01,menicucci14,chuang97,ketterer16,binomial16}.
Although each encoding has its respective operational advantage, 
the existing encodings have one property in common: the logical bases are pure physical states.
Therefore, logical basis initialisation necessarily requires a pure qumode state to be prepared from the equilibrium, i.e. thermal state.  
This requirement can be fulfilled if the physical system involves negligible thermal excitation, e.g. optical mode \cite{CVMBQC}, or if ground-state cooling is efficient, e.g. motional state of trapped ions \cite{Leibfried:2003tg}.  However, there are also bosonic systems that thermal excitation is significant, but ground-state cooling is challenging, e.g. mechanical oscillator \cite{Poot:2012fh}, or requires demanding refrigerating facilities \cite{book:cold, OConnell:2010br, 2017NatCo...815189T}.  If a pure-state encoding is employed in these systems, the physical impurity will contaminate the encoded logical information.

Fortunately, purifying a qumode is sufficient but not necessary for logical basis initialisation; recently is has been found that a \textit{pure} logical state can be encoded by a \textit{mixed} physical state \cite{kero16}.  While counter-intuitive at first glance, mixed-state encoding (MSE) shares a similar concept as noiseless subsystems in DV systems: quantum information is represented not only by a particular physical state, but by any state in a subspace of the Hilbert space \cite{Knill:2000wt, Zanardi:2001vo, Kempe:2001bg}.  

To the best of our knowledge, the idea of encoding qubits by highly-mixed CV states was first proposed by Jeong and Ralph \cite{jeong06,jeong07}.  By transferring a qubit information into differently displaced thermal states, the non-classical property of the qumode prevails even when the thermal excitation is arbitrarily high.  Unfortunately, the proposal did not discuss the explicit implementation of the logical operations which are required for universal quantum computation (UQC).  Recently, some of us introduced the two-qumode parity (TQP) encoding, which encodes each logical qubit by two mixed-state qumodes with opposite parities \cite{kero16}.  All UQC logical operations, including logical basis initialisation, logic gates, and information readout, can be implemented by realistic physical processes.  

In spite of certain advantages over pure-state encodings, the practical utility of TQP encoding is limited due to three major drawbacks.  First, each TQP qubit involves two qumodes, which increases the difficulty of implementation and squanders the information capacity provided.
Second, logical basis initialisation requires quantum nondemolition (QND) measurement and post-selection, which are challenging for many physical systems.  Third, the readout of quantum information requires fine-grained, DV parity measurement; in some platform this is less efficient than CV measurement, e.g. homodyne detection \cite{Hadfield:2009hd, 2015NatCo...6E6665F}.  

In this work, we propose a new MSE that does not suffer from the drawbacks of the TQP encoding.  
In this new encoding, the qubit computational value is represented by the parity of a single qumode, and the qubit coherence is represented by the sign of a quadrature of the qumode wavefunction.
An arbitrarily pure logical state can be initialised from equilibrium by deterministically displacing a physical thermal state.  
All UQC logic gates and information readout can be implemented by realistic physical processes and homodyne detection.  
In principle, our encoding eliminates the necessity of ground state cooling, so it can extend quantum computing candidacy to physical systems where cooling is inefficient.  
Furthermore, even for the physical platforms that cooling is efficient, our scheme allows quantum computation to be operated with more accessible refrigeration facility \cite{book:cold}, and can improve the noise tolerance of the encoded logical information. 

Our paper is organised as follows: a general formalism of MSE is presented in Sec.~\ref{sec:MSE}.  Our new encoding is introduced in Sec.~\ref{sec:QSP}.  The practical procedure for implementing UQC is also presented.  In Sec.~\ref{sec:FT}, we discuss about how fault-tolerance can be introduced through concatenating higher level error correction on top of our scheme.  
In Sec.~\ref{sec:NT}, we provide an explicit example that our scheme can improve the noise tolerance of a pure-state encoded qubit.  A conclusion is given in Sec.~\ref{sec:conclusion}.

\section{Pure- and mixed-state encoding \label{sec:MSE}}

In pure-state encodings, two pure, orthogonal, physical CV states, $\ket{\psi_0}$ and $\ket{\psi_1}$, are assigned as logical bases to represent the computational values ``0" and ``1".  Examples of such pure-state basis include Fock states, coherent states, cat states, and else  \cite{chuang95, chuang96, knill01, ralph03, lund08, An09, leghtas13, mirrahimi14, gottesman01, menicucci14, chuang97, ketterer16, binomial16}.  
To initialise a logical qubit, the encoding physical system is prepared in a pure physical state within the computational subspace spanned by $\{\ket{\psi_0}, \ket{\psi_1} \}$.
Logic gates are physical operations that transform a state within the computational subspace,
or generally within a tensor product of such subspace that represents a multi-qubit state.  A logical readout can be implemented by a physical measurement that distinguishes $\ket{\psi_0}$ from $\ket{\psi_1} $.

In MSE, the computational values are no longer represented by two particular physical states, but by two subspaces $\{\ket{\psi_0^{(1)}},\ket{\psi_0^{(2)}},\ldots \}$ and $\{\ket{\psi_1^{(1)}},\ket{\psi_1^{(2)}},\ldots \}$.  Each state in the subspaces are orthogonal, i.e. $\langle\psi_i^{(l)} | \psi_j^{(k)} \rangle = \delta_{ij}\delta_{kl}$, where $i,j\in\{0,1\}$ is the computational value; $k,l$ is the index of basis state in each subspace.

At initialisation, the physical system has to be prepared in a pure logical state.  For each logical qubit, this could be a physical state that represents,
e.g. a logical computational basis $\ket{0_L}$ or $\ket{1_L}$, or a logical coherence basis $\ket{+_L}$ or $\ket{-_L}$.
In contrast to pure-state encoding, MSE does not require a pure logical state to be represented by a pure physical state.  For example, to initialise $\ket{0_L}$,
the encoding qumode can be prepared in any (pure or mixed) physical state within the subspace $\{\ket{\psi_0^{(k)}} \}$.  

After initialisation, physical operation $\hat{U}$ is applied to implement logic gates.  A MSE logic gate is required to transform every state in each subspace in the same fashion \cite{kero16}.  Explicitly, for any multi-qubit basis $|\psi_{i_1}^{(k_1)}\psi_{i_2}^{(k_2)}\ldots \rangle$, the amplitude of transformation $\tilde{c}_{i_1i_2\ldots, j_1 j_2 \ldots}$ has to be independent of the basis index $\{k_1,k_2,\ldots\}$, i.e.
\begin{equation}
\hat{U}\left|\psi_{i_1}^{(k_1)}\psi_{i_2}^{(k_2)}\ldots \right\rangle = \sum_{j_1, j_2, \ldots} \tilde{c}_{i_1i_2\ldots, j_1 j_2 \ldots} \left|\psi_{j_1}^{(k_1)}\psi_{j_2}^{(k_2)}\ldots \right\rangle~,
\end{equation}
where $i_n$ and $j_n$ are the computational value of the $n$th qubit; $k_n$ is the basis index of the $n$th qubit.  After computation, quantum information can be read out by a physical measurement that distinguishes the subspaces  $\{\ket{\psi_0^{(k)}} \}$ and $\{\ket{\psi_1^{(k)}} \}$.

To see how this formalism permits quantum computation with mixed physical states, we note that the computational result is determined by the probability of each logical measurement outcome, $|\tilde{c}_{i_1i_2\ldots, j_1 j_2 \ldots}|^2$, which is independent of the basis index.  
Therefore, the same computational result is generated even if the initial physical state is a mixture of 
basis states with different basis indices.

Following the spirit of UQC, any logical unitary transformation can be decoupled into a sequence of logic gates \cite{book:NielsenChuang}.  Ref.~\cite{kero16} shows that any (single- and multi-qubit) MSE logic gate can be sufficiently generated by exponentiating the tensor product of the physical operators 
\begin{equation}\label{eq:XZ_decompose}
\hat{X}_E \equiv \sum_k \hat{X}^{(k)}~~\textrm{and}~~\hat{Z}_E \equiv \sum_k \hat{Z}^{(k)}~,
\end{equation}
where $\hat{X}^{(k)}$ and $\hat{Z}^{(k)}$ are the Pauli operators of the $k$th pair of basis $\{\ket{\psi_0^{(k)}},\ket{\psi_1^{(k)}} \}$  (see definition in Appendix~\ref{app:Pauli}).  The subscript $E$ denotes the operator is acting on physical states.  
Within the encoding subspace, these operators satisfy
\begin{equation}\label{eq:XZ}
\hat{X}_E^2 = \hat{Z}_E^2 = \hat{I}_E~,~\left[\hat{X}_E,\hat{Z}_E \right]_+ =0~,
\end{equation}
where $\hat{I}_E \equiv \sum_n \hat{I}^{(n)}$ is the identity operator;  $[\cdot, \cdot]_+$ is the anti-commutator.  We also define 
\begin{equation}\label{eq:Y}
\hat{Y}_E \equiv i \hat{X}_E \hat{Z}_E= \sum_k \hat{Y}^{(k)}~.
\end{equation}

It is easy to see that these physical operators obey the same algebra as Pauli matrices, i.e., for $\mu,\nu \in \{1,2,3 \}$
\begin{subequations}
\begin{eqnarray}\label{eq:APO_commute}
\left[\hat{\mathcal{Q}}_{\mu,E}, \hat{\mathcal{Q}}_{\nu,E} \right] &=& 2i \epsilon_{\mu \nu \omega}\hat{\mathcal{Q}}_{\omega,E}~,  \\
\left[\hat{\mathcal{Q}}_{\mu,E}, \hat{\mathcal{Q}}_{\nu,E} \right]_+&=& 2\delta_{\mu \nu} \hat{\mathcal{Q}}_{0,E} ~,\label{eq:APO_anticommute}
\end{eqnarray}
\end{subequations}
where $\{\hat{\mathcal{Q}}_{0},\hat{\mathcal{Q}}_{1},\hat{\mathcal{Q}}_{2}, \hat{\mathcal{Q}}_{3} \}\equiv \{\hat{I}, \hat{X}, \hat{Y}, \hat{Z} \}$.  We hereafter refer these physical operators as \textit{analogous Pauli operators} (APO).  

Before moving forward, we briefly discuss the physical meaning of the definitive APO, $\hat{Z}_E$ and $\hat{X}_E$.  $\hat{Z}_E$ classifies the physical states into basis subspaces that represent different computational values.  In fact, such classification is also employed in classical computation: a logical value is usually encoded by a physical state that is not fully characterised.  
For instance, a light bulb can represent two bit values by two sufficiently distinct brightnesses; 
within a reasonable range, a fluctuation of brightness will not affect the encoded bit value.

The crucial difference between classical and quantum computation is that the latter permits a coherent superposition of computational values.  In pure-state encoding, this is represented by a superposition of pure physical basis states.  In MSE, however, a ``coherent superposition of mixed state" does not make sense.  We recall that quantum superposition arises because not all quantum operators commute.  In pure-state encodings, a coherent superposition could be characterised by an operator that does not commute with Pauli $Z$, such as Pauli $X$.  MSE generalises this idea to all basis pair in the subspace (c.f. Eq.~(\ref{eq:XZ_decompose})).  In other words, $\hat{X}_E$, which does not commute with $\hat{Z}_E$, characterizes the coherence of a MSE qubit.  We note that a MSE qubit with coherent superposition could then be recognized as a ``mixture of pure superposition states".

\subsection{Representation of logical information \label{ssec:representation}}

As adopted from Ref.~\cite{kero16}, Eq.~(\ref{eq:XZ_decompose}) expresses an APO as a summation of Pauli operators of each basis pair.  This definition could intuitively explain the key idea of MSE, i.e. physical purity is not necessary for UQC.  However, if we want to use this definition to evaluate the quantum information encoded in a general physical state, the state has to be resolved into each of the basis states.  Because a qumode could exhibit infinite basis pairs, the evaluation process is generally tedious.  

Alternatively, we introduce another definition of APO: by the physical Hermitian operators that obey the algebra in Eqs.~(\ref{eq:APO_commute}) and (\ref{eq:APO_anticommute}).  The Hermitian operators could be expressed in terms of the qumode operators ($\hat{a}$ and $\hat{a}^\dag$), without resolving into the pure-state bases (though it could, see Appendix~\ref{app:equivalence}).  An immediate advantage is that the physical implementation of the logic gates can be more easily inferred as physical interaction is usually described in terms of qumode operators but not basis states.

For any MSE, a logical qubit can be expressed by the APO as
\begin{equation}\label{eq:logical_qubit}
\rho_L = \frac{1}{2}\left(\langle \hat{I}_E \rangle \hat{I}_L + \langle \hat{X}_E \rangle \hat{X}_L + \langle \hat{Y}_E\rangle \hat{Y}_L + \langle \hat{Z}_E \rangle \hat{Z}_L \right)~.
\end{equation}
$\hat{\mathcal{Q}}_L$'s are the Pauli operators for the logical basis states $\{\ket{0_L}, \ket{1_L} \}$.  We can see another advantage of this definition: the encoded quantum information can be evaluated by simply calculating the expectation values, $\langle \hat{\mathcal{Q}}_E \rangle \equiv \textrm{Tr}\{\hat{\mathcal{Q}}_E \rho \}$ for any physical state $\rho$.  This is particularly useful in analysing the performance of MSE encoded qubit under noisy processes.

Similarly, an $N$-qubit logical state can be represented by
\begin{equation}\label{eq:N_partite}
\rho_L = \frac{1}{2^N} \sum_{\bm{\mu}} \Tr{ \hat{\mathcal{Q}}_{\bm{\mu},E } \rho } \hat{\mathcal{Q}}_{\bm{\mu},L}~,
\end{equation}
where $\bm{\mu}\equiv \{\mu_1, \mu_2, \ldots \}$ for $\mu_n \in \{0,1,2,3\}$; $\hat{\mathcal{Q}}_{\bm{\mu}} \equiv \hat{\mathcal{Q}}_{\mu_1}\otimes \hat{\mathcal{Q}}_{\mu_2}\otimes \ldots \hat{\mathcal{Q}}_{\mu_N}$.  We note that the number of qubits is not necessarily the same as the number of qumodes because each qubit can be encoded by multiple qumodes, i.e. $\hat{\mathcal{Q}}_{\mu_i}$ can be multi-mode operator.

We emphasise that the logical state $\rho_L$ resides in the hypothetical logical Hilbert space, which should not be confused with the physical Hilbert space in which the qumode state $\rho$ resides.
In general, different physical states can produce the same expectation values $\langle \hat{\mathcal{Q}}_E\rangle$, so the same logical state $\rho_L$ can be represented by either one or a mixture of such physical states.  Therefore, the purity of $\rho$ is generally not the same as $\rho_L$.  
In fact, a pure logical qubit can be encoded by an arbitrarily mixed physical state \cite{jeong06, jeong07, kero16}.

\subsection{Logic gates}

A general $N$-qubit logical state can be expressed in terms of the logical operators, i.e.
\begin{equation}
\rho_L = \frac{1}{2^N}\sum_{\bm{\mu}} \Tr{\hat{\mathcal{Q}}_{\bm{\mu},L} \rho_L} \hat{\mathcal{Q}}_{\bm{\mu},L}~.
\end{equation}
In quantum computation, the quantum algorithm is specified by a unitary transformation $\hat{U}_L$, which transforms the logical state as 
\begin{eqnarray}\label{eq:transformed_rhoL}
\hat{U}_L \rho_L \hat{U}_L^\dag &=& \frac{1}{2^N} \sum_{\bm{\mu}} \Tr{\hat{\mathcal{Q}}_{\bm{\mu},L} \hat{U}_L\rho_L \hat{U}_L^\dag} \hat{\mathcal{Q}}_{\bm{\mu},L} \nonumber \\
& = &  \frac{1}{2^N} \sum_{\bm{\mu} \bm{\mu}'} c_{\bm{\mu} \bm{\mu}'} \Tr{\hat{\mathcal{Q}}_{\bm{\mu}',L}  \rho_L } \hat{\mathcal{Q}}_{\bm{\mu},L} ~.
\end{eqnarray}
The operation of $\hat{U}_L$ is defined by its transformation coefficients $c_{\bm{\mu} \bm{\mu}'}$ of each logical operator, i.e.
\begin{equation}
\hat{U}^\dag_L \hat{\mathcal{Q}}_{\bm{\mu},L} \hat{U}_L \equiv \sum_{\bm{\mu}'} c_{\bm{\mu} \bm{\mu}'} \hat{\mathcal{Q}}_{\bm{\mu}',L} ~.
\end{equation}
By the virtue of UQC, any logical unitary can be constructed by applying a universal set of logic gates in an appropriate sequence \cite{book:NielsenChuang}, i.e.
\begin{equation}
\hat{U}_L = \mathcal{U}(\hat{\mathcal{Q}}_{\bm{\mu},L}) = u_1(\hat{\mathcal{Q}}_{\bm{\mu},L}) u_2(\hat{\mathcal{Q}}_{\bm{\mu},L})\ldots ~,
\end{equation}
where $\mathcal{U}$ and $u$ are functional of Pauli operators; the subscript of $u$ denotes the sequence of logic gates.  Depending on the choice of universal gate set, each $u$ can be composed of the Pauli operators of at most two qubits.  One such choice is \cite{kero16}
\begin{equation}\label{eq:gate_set}
u\in \{ e^{i\theta \hat{X}},  e^{i\phi \hat{Z}},  e^{i\phi \hat{Z}\otimes \hat{Z}}\}~,
\end{equation}
where $\theta$ and $\phi$ are controllable real numbers.  The first two operations correspond to the single-qubit $X$- and $Z$-axis rotation, which suffice to implement any single-qubit unitary.  The last operation is a two-qubit conditional $Z$ rotation, which generates entanglement. 

For any encoded state given by Eq.~(\ref{eq:N_partite}), a physical transformation $\hat{U}_E$ can implement the logical unitary $\hat{U}_L$ if the transformed physical state $\hat{U}_E \rho \hat{U}_E^\dag$ encodes the logical state $\hat{U}_L \rho_L \hat{U}_L^\dag$ in Eq.~(\ref{eq:transformed_rhoL}), i.e.
\begin{equation}
\frac{1}{2^N} \sum_{\bm{\mu}} \Tr{\hat{\mathcal{Q}}_{\bm{\mu},E} \hat{U}_E \rho \hat{U}_E^\dag} \hat{\mathcal{Q}}_{\bm{\mu},L} = \hat{U}_L \rho_L \hat{U}_L^\dag ~.
\end{equation}
It is easy to see that this criterion is satisfied if the physical transformation obeys
\begin{equation}
\hat{U}^\dag_E \hat{\mathcal{Q}}_{\bm{\mu},E} \hat{U}_E \equiv \sum_{\bm{\mu}'} c_{\bm{\mu} \bm{\mu}'} \hat{\mathcal{Q}}_{\bm{\mu}',E} ~,
\end{equation}
 for any $\bm{\mu}$.  Because the APO follow the same algebra as Pauli operators, $\hat{U}_E$ can be constructed as
\begin{equation}\label{eq:U_physical}
\hat{U}_E \equiv \mathcal{U}(\hat{\mathcal{Q}}_{\bm{\mu},E})=u_1(\hat{\mathcal{Q}}_{\bm{\mu},E}) u_2(\hat{\mathcal{Q}}_{\bm{\mu},E})\ldots~.
\end{equation} 
Hence UQC can be implemented by realising the basic physical operations $u(\hat{\mathcal{Q}}_{\bm{\mu},E})$, e.g. those in Eq.~(\ref{eq:gate_set}), which act as analogous logic gates.

\subsection{Projective measurement \label{ssec:MSE_Eeasurement}}

Apart from unitary transformation, projective measurement is another important logical operation.  Two main utilities of projective measurement are: to extract the processed quantum information,
and to post-selectively apply a projection to the unmeasured qubits.

Any single-qubit projective measurement is equivalent to a Pauli basis measurement after qubit rotation \cite{book:Mosca}.  Without loss of generality, we consider the $X$-basis logical qubit measurement.  If the first logical qubit of the $N$-qubit state in Eq.~(\ref{eq:N_partite}) is measured in $\hat{X}_L$ basis, the remaining $(N-1)$-qubit (unnormalised) state becomes
\begin{eqnarray}\label{eq:logical_Measure}
& & \bra{\pm_{L_1}}\rho_L \ket{\pm_{L_1}}  \\
&=& \frac{1}{2^{N-1}} \sum_{\bm{\mu} \backslash \mu_1} \Tr{\Big(\frac{\hat{I}_{E_1} \pm \hat{X}_{E_1}}{2} \otimes  \hat{\mathcal{Q}}_{\bm{\mu},E}\Big) \rho }  \hat{\mathcal{Q}}_{\bm{\mu},L} \nonumber \\
&=& \frac{1}{2^{N-1}} \sum_{\bm{\mu} \backslash \mu_1} \textrm{Tr}_{N\backslash 1}\left\{ \hat{\mathcal{Q}}_{\bm{\mu},E} \textrm{Tr}_1\Big\{\Big(\frac{\hat{I}_{E_1} \pm \hat{X}_{E_1}}{2}\Big) \rho \Big\} \right\}  \hat{\mathcal{Q}}_{\bm{\mu},L} \nonumber
~.
\end{eqnarray}
We note that the subscript $j$ of $L_j$ and $E_j$ indicates the $j$th logical qubit and the qumode(s) representing it.  The normalisation of the above state is the probability $\mathbb{P}$ of obtaining the outcome $\pm$, i.e.
\begin{equation}
\mathbb{P}(\pm) =  \Tr{\bra{\pm_{L_1}}\rho_L \ket{\pm_{L_1}}} = \Tr{\frac{\hat{I}_{E_1} \pm \hat{X}_{E_1}}{2}\rho}~.
\end{equation}

In Eq.~(\ref{eq:logical_Measure}), we can see that the unnormalised physical state, which represents the projected $(N-1)$-qubit state, can be produced by projecting the qubit-1 qumodes with the physical projectors $(\hat{I}_E \pm \hat{X}_E)/2$.  As a result, physically realising these projectors implement the logical $X$-basis measurement in MSE.

\subsection{Summary for mixed-state encoding}

We have shown that if there are two physical operators $\hat{X}_E$ and $\hat{Z}_E$ that obey Eq.~(\ref{eq:XZ}), a set of APO can be constructed that obeys the same algebra as Pauli operators, Eqs.~(\ref{eq:APO_commute}) and (\ref{eq:APO_anticommute}).  These APO specify all essential ingredients of an encoding: how a multi-qubit logical state is represented (Eq.~(\ref{eq:N_partite})), what physical operations are required to implement the universal logic gates (Eq.~(\ref{eq:U_physical})), and what physical projectors can implement the logical projective measurement (Eq.~(\ref{eq:logical_Measure})).  From Eq.~(\ref{eq:N_partite}), we can see that different physical states can represent the same logical state if they have the same expectation values for all APO.  This permits a pure logical state to be encoded by a mixed physical state.

\section{Quadrature-sign parity encoding \label{sec:QSP}}

So far, the TQP encoding in Ref.~\cite{kero16} is the only known MSE that all logical operations for UQC are specified.  In this encoding, a logical qubit is encoded by two qumodes with opposite parity.  The $\hat{X}_E$ and $\hat{Z}_E$ of this encoding are respectively the two-qumode symmetry and the parity of the second qumode.  Although the TQP encoding could allow UQC without ground state cooling, its practical utility is limited due to three main drawbacks: two qumodes are required to encode a qubit, measurement and post-processing are required to initialise a logical basis, and information readout requires DV parity measurement.

\begin{figure}
\begin{center}
\includegraphics[width=\linewidth]{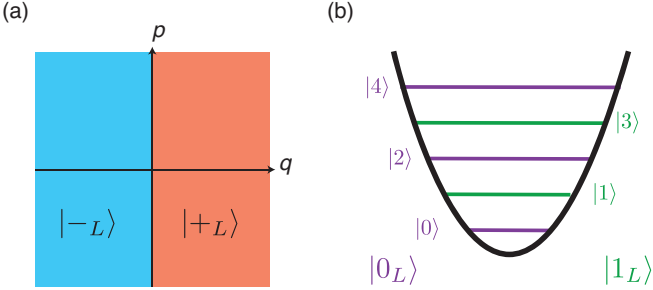}
\caption{ Physical basis states corresponding to the QSP encoding APO in Eq.~(\ref{eq:XZ}).
(a) Every state where the Wigner function contains only positive (negative) $q$-quadrature variable can encode the logical coherence basis $\ket{+_L}$ ($\ket{-_L}$).  
(b) Every state composed of only even (odd) boson number can encode the logical computational basis $\ket{0_L}$ ($\ket{1_L}$).
\label{fig:QSP}} 
\end{center}
\end{figure}

We here present a new encoding that does not suffer from any of these drawbacks.  The definitive APO are given by
\begin{equation}\label{eq:XZ_QSP}
\hat{X}_E \equiv \int_{-\infty}^\infty \Theta(x) \ket{x_q}\bra{x_q} dx ~~,~~\hat{Z}_E \equiv e^{i\pi \hat{a}^\dag \hat{a}} = \hat{\mathcal{P}}~,
\end{equation}
where $\Theta(x)$ is the sign function; $\hat{\mathcal{P}}$ is the parity operator; $\ket{x_q}$ is the $q$-quadrature eigenstate with eigenvalue $x$; the quadratures follow the standard definition $\hat{a}\equiv (\hat{q}+i\hat{p})/\sqrt{2}$.  A schematic illustration of the corresponding basis subspace is shown in Fig.~\ref{fig:QSP}.  It is straightforward to check that these operators obey Eq.~(\ref{eq:XZ}), where the encoding space is the entire space of a qumode:
\begin{equation}\label{eq:I_QSP}
\hat{I}_E = \int_{-\infty}^\infty \ket{x_q}\bra{x_q}dx~.
\end{equation}
The remaining APO, $\hat{Y}_E$, is defined by Eq.~(\ref{eq:Y}).

We hereby call our new encoding quadrature-sign parity (QSP) encoding.  Because all APO are single-mode operators, each QSP qubit is encoded by only one qumode.  
This circumvents the TQP encoding drawback that requires two qumodes per qubit.

\subsection{Logical basis initialisation \label{ssec:initialise}}

As seen from the definition of $\hat{X}_E$ in Eq.~(\ref{eq:XZ_QSP}), any state of which the wave function has non-vanishing amplitude only at positive (negative) $q$-quadrature, i.e. its Wigner function resides in the right (left) half of phase space, can encode the logical coherence basis $\ket{+_L}$ ($\ket{-_L}$).  
This feature allows a deterministic initialisation of logical basis from physical equilibrium (i.e. thermal state) by applying unitary displacement in $q$-quadrature.  As such, initialising a QSP qubit is more efficient than initialising a TQP qubit, which requires QND measurement and post-selection \cite{kero16}.

Quantitatively, a displaced thermal state with real and positive displacement $\alpha$, i.e. $\rho = \rho_D \equiv \hat{D}(\alpha) \rho_\textrm{th} \hat{D}^\dag(\alpha)$, exhibits logical infidelity to the ideal $\ket{+_L}$ as
\begin{eqnarray}\label{eq:logical_basis_fidelity}
1-\bra{+_L} \rho_L \ket{+_L} &=& \frac{1}{2}\left(1- \Tr{\hat{X}_E \hat{D}(\alpha)\rho_\textrm{th}\hat{D}^\dag(\alpha)}\right) \nonumber \\
&=& \frac{1}{2}\textrm{erfc} \left(\frac{\alpha}{\sqrt{\bar{n}+1/2}} \right)~,
\end{eqnarray}
where $\hat{D}(\alpha) \equiv \exp\left(\alpha \hat{a}^\dag-\alpha^\ast \hat{a} \right)$ is the displacement operator; $\rho_\textrm{th}$ is the physical thermal state with mean excitation $\bar{n} \equiv \Tr{\hat{a}^\dag \hat{a}\rho_\textrm{th}}$ \cite{weedbrook12}; $\textrm{erfc}(x)$ is the complementary error function.  For any $\bar{n}$, the logical infidelity can be reduced exponentially by increasing the displacement $\alpha$.  

We note that the representation of logical qubits by displaced thermal states may look similar to  \cite{jeong06,jeong07}.  However, we emphasize two crucial differences in our scheme.  First, most techniques presented in \cite{jeong06,jeong07} are dedicated to displaced thermal states; their applicability to other mixed state is likely but not discussed in detail.  In contrast, our scheme is applicable to any mixed state $\rho$ with confined $q$-quadrature variance, i.e., $1- \Tr{\hat{X}_E \hat{D}(\alpha)\rho \hat{D}^\dag(\alpha)} \ll 1$ for a sufficiently large $\alpha$.  
This feature of QSP encoding is useful for the quantum computer architecture that reuses qumodes after measurement, where the after-measurement state is generally not thermal state.  Because displacement operation is usually faster than dissipative cooling, QSP encoding can thus reduce the qubit reinitialisation time.

Second, \cite{jeong06,jeong07} focus on verifying the quantum properties of displaced thermal states, but the implementation of UQC logical operations is not explicitly discussed.  In the following, we will present the general procedure to quantum-compute with QSP encoding states.  This is the main contribution of the current work.

\subsection{Logic gate and qubit measurement}

UQC can be implemented by applying, in sequence, the basic operations in Eq.~(\ref{eq:gate_set}), which include analogous $X$-axis rotation $\exp(i\theta \hat{X}_E)$, analogous $Z$-axis rotation $\exp(i\theta \hat{Z}_E)$, and an entangling gate $\exp(i\theta \hat{Z}_E\otimes \hat{Z}_E)$.  
In QSP encoding, the latter two are respectively single- and two-qumode exponential parity gates:
\begin{equation}
e^{i\theta \hat{Z}_{E_j}} = e^{i \theta \hat{\mathcal{P}}_j}\equiv \hat{R}_j(\theta)~~,~~
e^{i \theta \hat{Z}_{E_j}  \hat{Z}_{E_l}} = 
e^{i \theta \hat{\mathcal{P}}_j \hat{\mathcal{P}}_l} \equiv \hat{\mathcal{E}}_{jl}(\theta)~,
\end{equation}
where $\hat{\mathcal{P}}_i$ is the parity operator of the $i$-th qumode.  These gates can be deterministically implemented by dispersively coupling the qumodes to an auxiliary physical qubit \cite{kero16-2, Lau:2017ky, kero16}.  Its experimental realisation has recently been demonstrated with superconducting microwave cavities \cite{eswapexp}.

Alternatively, the exponential-parity gate might also be realisable by the approach of universal CV quantum computation \cite{Lloyd:1999vz, Sefi:2011df}.  In this approach, any Hamiltonian consisting of a polynomial order of quadrature operators can be efficiently engineered by concatenating lower order Hamiltonians.  
At first glance, this approach is deemed not applicable to the exponential-parity gate, because the series expansion of its effective Hamiltonian, $\hat{\mathcal{P}}$, involves an infinite order of quadrature operators,
\begin{equation}\label{eq:Parity_qexpansion}
\hat{\mathcal{P}} = \sum_{k=0}^\infty \frac{(i \pi)^k}{k!} \left(\hat{a}^\dag\hat{a} \right)^k =  \sum_{k=0}^\infty \frac{(i \pi)^k}{k!} \left( \frac{\hat{q}^2+ \hat{p}^2 -1}{2} \right)^k~.
\end{equation}

Nevertheless, for our purpose it is not necessary to implement an exponential-parity gate that is accurate for any state.  We would be satisfied if the gate is accurate with respect to our QSP qubit, which is initialised as a displaced thermal state.  We note that because all terms in Eq.~(\ref{eq:Parity_qexpansion}) preserve boson number, if the Hamiltonian is implemented accurately the processed physical states will share the same boson number distribution as the displaced thermal state.  
For these states, the population of high boson number decreases exponentially. 
Specifically, we show in Appendix~\ref{app:e-parity} that the population with boson number $r_{\max} \gtrsim 30 (\bar{n}+1/2)$ is negligible.  
The series in Eq.~(\ref{eq:Parity_qexpansion}) can then be truncated at a finite order $k_{\max}$ without introducing significant gate error (details in Appendix~\ref{app:e-parity}).  

The remaining operation is $\exp(i\theta \hat{X}_E)$, which is difficult to physically implement because $\hat{X}_E$ is highly nonlinear.  Nevertheless, the necessity of this gate can be circumvented by employing, instead of the circuit-based model, measurement-based quantum computation (MBQC) \cite{raussendorf01, raussendorf03,briegel09}.  

Implementing UQC with MBQC involves two criteria: first, the ability to prepare a cluster state with a specific graph; second, logical qubit measurement in any basis on the $X$-$Y$ plane.  In the following, we will show that both criteria can be deterministically implemented with QSP encoding.

\subsubsection{Cluster state construction \label{sssec:cluster}}

A cluster state is prepared by applying logical controlled-phase gate (CPhase, see Appendix~\ref{app:cphase} for the logic table) to qubits that are initialised as $\ket{+_L}$.  The quantum computing algorithm executed by MBQC is determined by the graph of the cluster state.  Each vertex of the graph denotes a logical qubit, and each edge specifies the qubits that have to be entangled by CPhase.  Such cluster state with a specific structure is usually referred as a graph state.

To prepare a graph state in QSP encoding, all qumodes are first initialised as $\ket{+_L}$ by displacement (c.f. Sec.~\ref{ssec:initialise}).  
The logical CPhase gate can be deterministically implemented by a sequence of exponential-parity gates, i.e.
\begin{equation}\label{eq:CPhase}
\hat{\mathcal{C}}_{jl}\equiv e^{i\frac{\pi}{4}(\hat{I}_E-\hat{Z}_E)_j (\hat{I}_E-\hat{Z}_E)_l} = 
\hat{R}_j(-\frac{\pi}{4})\hat{R}_l(-\frac{\pi}{4}) \hat{\mathcal{E}}_{jl}(\frac{\pi}{4})~,
\end{equation}
where an unimportant global phase is omitted.  A graph state can be deterministically prepared by applying CPhase to the qumodes that represent the edge-connected vertices.

We note that in MBQC literatures \cite{raussendorf01, raussendorf03,briegel09}, a graph state is usually constructed by first preparing a 2D cluster state, then a collection of qubits is distangled by $Z$-basis measurement.  
However, $Z$-basis measurement is not necessary if the CPhase gate can be deterministically applied to selected qubits. 
This could be advantageous to the physical platforms where a $Z$-basis measurement (i.e. parity measurement in QSP encoding) is challenging to realise.

\subsubsection{Logical $X$-axis measurement \label{sssec:logxmeas}}

As discussed in Sec.~\ref{ssec:MSE_Eeasurement}, a logical $X$-axis measurement can be realised by a physical measurement with projectors $(\hat{I}_E \pm \hat{X}_E)/2$. Because $\hat{X}_E$ in the QSP encoding is the sign of $q$-quadrature, intuitively its projector could be implemented by homodyne-detecting the qumode in $q$-quadrature, and distinguishing the outcome by its sign.  

To verify this intuition, we consider when the first qumode of an $N$-qumode state is homodyne-detected in $q$-quadrature.  For an measurement outcome $x$, the remaining $(N-1)$-qumode state is projected to
\begin{equation}
\rho'(x) \equiv \frac{\bra{x_q}\rho \ket{x_q}}{\Tr{\bra{x_q}\rho \ket{x_q}}}~.
\end{equation}
The probability of obtaining an outcome between $x$ and $x +dx$ is
\begin{equation}
\mathbb{P}(x) dx = \Tr{\bra{x_q}\rho \ket{x_q}} dx~.
\end{equation}

If we retain no information but the sign ($\pm$) of the outcome, the remaining $(N-1)$-qumode state is conditionally projected to
\begin{eqnarray}
\pm \int_0^{\pm \infty} \mathbb{P}(x) \rho'(x) dx &=& \pm \int_0^{\pm \infty} \bra{x_q} \rho \ket{x_q} dx\nonumber \\
& = &  \textrm{Tr}_1 \left\{ \Big( \pm \int_0^{\pm \infty} \ket{x_q}\bra{x_q}dx \Big) \rho  \right\} \nonumber \\
& = & \textrm{Tr}_1 \left\{ \frac{\hat{I}_{E_1} \pm \hat{X}_{E_1}}{2} \rho  \right\}~.
\end{eqnarray}
The last relation shows that the physical projector $(\hat{I}_E \pm \hat{X}_E)/2$ is implemented on qumode 1.  Hence the logical $X$-basis measurement of QSP encoding is realised.

\subsubsection{Logical $X$-$Y$ plane and $Z$-axis measurement \label{sssec:logzmeas}}

We remind that a measurement along any axis on the $X$-$Y$ plane, i.e. $(\cos\theta \hat{X} + \sin \theta \hat{Y})$ for any $\theta$, can be implemented by applying a $Z$-axis rotation before a $X$-basis measurement \cite{book:Mosca}, i.e.
\begin{eqnarray}
&&\Tr{\frac{\hat{I}_L \pm (\cos\theta \hat{X}_L + \sin\theta \hat{Y}_L)}{2}\rho_L } \nonumber \\
&=& \Tr{\frac{\hat{I}_L \pm \hat{X}_L}{2} e^{i\frac{\theta}{2}\hat{Z}_L}\rho_L e^{-i\frac{\theta}{2}\hat{Z}_L}}~.
\end{eqnarray}
Following this idea, a logical $X$-$Y$ plane measurement in the QSP encoding can be realised by first 
applying $Z$-axis rotation, i.e. $\rho \rightarrow \hat{R}(\theta/2) \rho \hat{R}^\dag(\theta/2)$, then the qumode is measured in the logical $X$-basis (c.f. Sec.~\ref{sssec:logxmeas}).

At the end of MBQC, quantum information is typically read out in the $Z$-basis.  A $Z$-basis measurement can be physically implemented by the projectors $(\hat{\mathbb{I}}\pm \hat{Z}_E)/2$, which is the parity measurement in QSP encoding.  Nevertheless, physically realising $Z$-basis measurement is sufficient but not necessary.  Alternatively, we can always modify our quantum computing algorithm to execute an extra round of Hadamard gate on each result qubit.  
Because a $X$-basis measurement is equivalent to a $Z$-basis measurement after a Hadamard gate \cite{book:Mosca}, the quantum information can then be read out by $X$-basis measurement.

\begin{figure}
\begin{center}
\includegraphics[width=\linewidth]{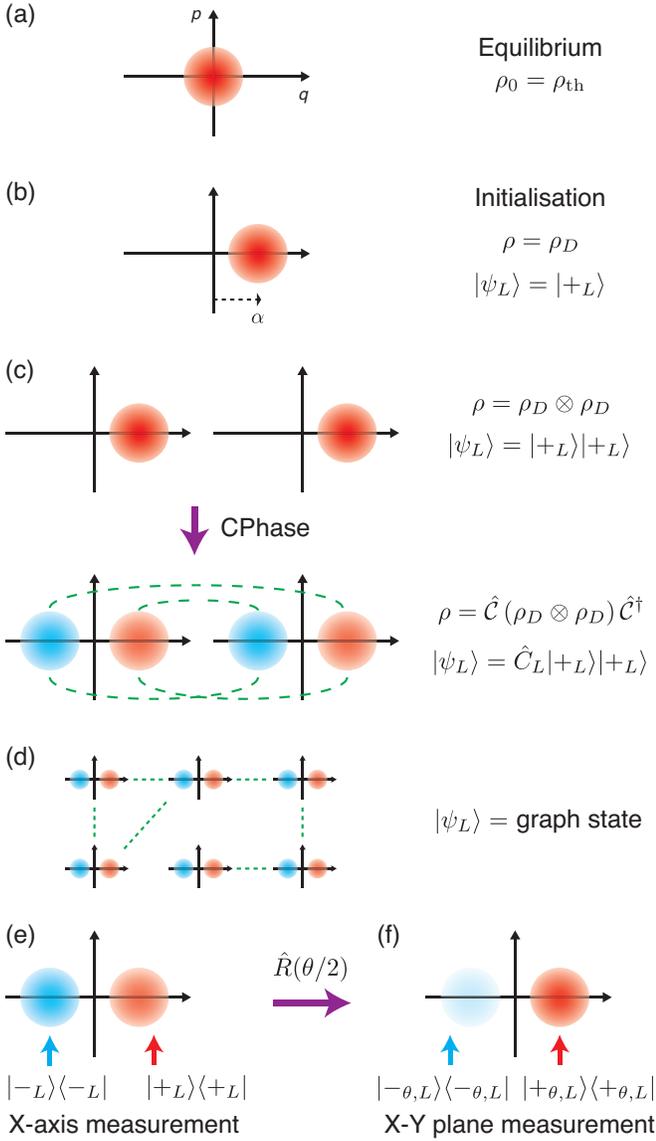}
\caption{Procedure for quantum computing with QSP encoding.  At each stage, the Wigner function of qumode is illustrated.  $\rho$ and $\ket{\psi_L}$ respectively denote the physical and logical state.  (a)  Before computation, each qumode is in thermal state $\rho_\textrm{th}$.  (b)  Logical coherence basis is deterministically initialised by displacing the thermal state, i.e. $\rho=\rho_D=\hat{D}(\alpha) \rho_\textrm{th} \hat{D}^\dag(\alpha)$.  (c)  Two logical qubits are entangled by logical CPhase gate, which can be realised by a sequence of exponential-parity gates (c.f. Eq.~(\ref{eq:CPhase})).  (d) A graph state can be constructed by applying CPhase gates to selected qumodes.  (e) Logical $X$-basis measurement is implemented by $q$-quadrature homodyne detection.  The logical measurement outcome is determined by the sign of the homodyne detection outcome (denoted by arrows).  (f) Measurement along other basis on the logical $X$-$Y$ plane: logical $Z$-axis rotation is applied before logical $X$-axis measurement.  The logical state will be projected to $\ket{\pm_{\theta,L}} \equiv (\ket{0_L} \pm e^{i\theta} \ket{1_L})/\sqrt{2}$, which is the $\pm 1$ eigenstate of $(\cos\theta \hat{X}_L + \sin\theta \hat{Y}_L)$. \label{fig:procedure}} 
\end{center}
\end{figure}

The above physical processes complete the requirement of conducting universal MBQC with QSP encoding.  A summary of the procedure is shown in Fig.~\ref{fig:procedure}.

\section{Logical fault tolerance \label{sec:FT}}

So far we have discussed the ideal implementation of UQC with QSP encoding.  In practice, the protocol could suffer from various sources of error, such as imperfect initialisation due to finite displacement, faulty implementation of exponential-parity gates, decoherence of qumodes, etc.
These imperfection will lead to a faulty logical cluster state and inaccurate execution of the computing algorithm.

In principle, if the physical errors are sufficiently small, the encoded quantum information could be protected from faults by quantum error correction \cite{Campbell:2017bq, Babbush18, GKP_toric}.  
For our scheme that employs cluster-state MBQC, fault-tolerance can be introduced through concatenating an additional logical layer on top of the QSP logical cluster state.  In order to eliminate naming confusions, we call this additional layer the supra-logical layer (SLL).

Here we outline the procedure of how quantum error-correcting code is implemented through the SLL.  Due to its structural simplicity, we choose the topological code as presented in \cite{raussendorf2007topological} and \cite{fowler2009topological}. This code is equivalent to the surface code from \cite{fowler2012surface}, which forms the basis of the quantum computing architectures expected within the next few years.

The key idea of implementing fault-tolerant quantum computation is to execute the SLL quantum circuits in a fault-tolerant way.  For any quantum algorithm to be computed, its quantum circuit can be prepared for fault-tolerance in the following steps. First, the circuit is transformed into another one that has a well defined structure \cite{paler2017fault}: it consists of only SLL qubits initialised in a restricted set of states, interacting through SLL CNOT gates, and measured by SLL measurements. The second step is to compile the SLL circuit into topologically error-correcting structures that represent the initialisations, the CNOTs, and the measurements \cite{fowler2009topological}; the group of these structures is usually known as an assembly \cite{paler2017synthesis}.  
Each assembly structure has a shape (i.e. it is a 3D object), which specifies the graph of the lower level (QSP) logical cluster state and how computation is executed with such state.  Finally, the logical cluster state is constructed deterministically by the method in Sec.~\ref{sssec:cluster}.

After state preparation, SLL quantum computation is executed by performing specific measurements on the QSP logical qubits according to the rules of the assembly.  In the defects and braids encoding method from \cite{fowler2009topological, paler2017synthesis}, logical $Z$ measurements are employed for implanting the defects in the cluster state, and logical $X$ measurements are conducted for syndrome detection. Syndromes are used in the classical error correction algorithm that is running parallel to the quantum system. We note that the $Z$ measurements can be avoided in our scheme, by simply not entangling into the cluster the qubits which are known to be $Z$ measured later.

Apart from logical $X$ measurement and cluster state formation, an additional non-Clifford element is required for the universality of quantum computation.  In the topological error correction (surface code) scheme we are considering, this element is the preparation of SLL $T$-state, which is used to implement the non-Clifford $T$-gate \cite{Fowler:2009ep}.  A SLL $T$-state can be constructed from a logical qubit $T$-state:
\begin{equation}
\ket{T_L} \equiv \frac{1}{\sqrt{2}}(\ket{0_L} + e^{i\pi/4} \ket{1_L})~.
\end{equation}
In our QSP encoding, this state can be deterministically prepared by applying $Z$-axis rotation $\hat{R}(\frac{\pi}{8})$ to the logical $X$ basis $\ket{+_L}$.  In practice, this preparation is not-fault tolerant, which would introduce error in the SLL circuit.
Nevertheless, higher fidelity $T$-states can be distilled by consuming multiple copies of lower fidelity $T$-states \cite{fowler2012surface}.

In summary, quantum computation can be made fault-tolerant by concatenating a SLL on top of the QSP encoding logical cluster state. When using surface codes inside the SLL, fault-tolerance introduces only a computational resource overhead (hardware and time) dictated mainly by the chosen code distance and the $T$-state distillation.

\section{Noise tolerance \label{sec:NT}}

We have discussed the notion that the QSP encoding allows quantum computation to be implemented directly with thermal state qumodes.  This eliminates the necessity of cooling the qumodes to their ground state, which is thus an advantage in phyical systems where ground-state cooling is challenging or resources demanding.  On the other hand, for the systems that ground-state cooling is efficient, we now demonstrate that QSP encoding can also provide an advantage: improving the error tolerance of quantum information.  

It is known that every MSE exhibits noiseless subsystems (NS) \cite{Knill:2000wt, Zanardi:2001vo, Kempe:2001bg, kero16-2, kero16}.  When an encoding physical state is transformed by noise, the quantum information is not corrupted if the resultant state is within the same subspace.  This is in stark contrast to pure-state encodings, where quantum information is lost if the erroneous physical state is not composed of the encoding bases (unless error correction is executed \cite{2018PhRvA..97c2346A}).  

As a MSE, QSP encoding also exhibits certain NS.  To illustrate this idea, we show an explicit example where the QSP encoding can improve the dephasing tolerance of (pure-state) cat-code qubits.  The logical computational basis of cat code is given by \cite{ralph03}
\begin{equation}
\ket{0_\textrm{cs}} = \frac{1}{\mathcal{N}_+}(\ket{\alpha}+\ket{-\alpha})~~,~~\ket{1_\textrm{cs}} = \frac{1}{\mathcal{N}_-}(\ket{\alpha}-\ket{-\alpha})~,
\end{equation}
where $\mathcal{N}_\pm \equiv \sqrt{2(1 \pm \exp(-2 |\alpha|^2))}$; $\alpha$ is again real and positive.  The basis states have definite but opposite parity, so they are also logical computational basis of QSP encoding.  For sufficiently large $\alpha$, the logical $X$ basis are approximately coherent states, i.e. $\ket{\pm_\textrm{cs}} \approx \ket{\pm\alpha}$.  The Wigner function of these coherent states is localised in either side of the phase space, so these states are also logical coherence basis of QSP encoding.  
In fact, these are no coincidence: a displaced thermal state qubit will become a cat-code qubit when there is no initial thermal excitation, i.e. $\rho_\textrm{th}=\ket{0}\bra{0}$.

Under a pure dephasing process, a physical state $\rho$ evolves as \cite{book:Milburn}
\begin{equation}
\dot{\rho} = \kappa \left(\hat{a}^\dag\hat{a} \rho \hat{a}^\dag\hat{a} - \frac{1}{2}(\hat{a}^\dag\hat{a})^2 \rho -  \frac{1}{2} \rho (\hat{a}^\dag\hat{a})^2 \right) ~,
\end{equation}
where $\kappa$ is the dephasing rate.  For any initial state $\rho(0)$, the evolved state at time $t$ is given by
\begin{equation}
\rho(t) = \int_{-\infty}^{\infty} e^{-i \varphi \hat{a}^\dag \hat{a}} \rho(0) e^{i \varphi \hat{a}^\dag \hat{a}} \frac{1}{\sqrt{2 \pi \kappa t}}e^{-\frac{\varphi^2}{2 \kappa t}} d\varphi ~.
\end{equation}
$\rho(t)$ can be viewed as a statistical mixture of rotated initial state, where the rotation angle $\varphi$ follows a Gaussian distribution with variance $\overline{ \varphi^2} =\kappa t $ . 

Assume the initial physical state of a cat-code qubit is $\rho(0)=\ket{\theta,\phi}\bra{\theta,\phi}$, where
\begin{equation}
\ket{\theta,\phi} \equiv \cos\frac{\theta}{2} \ket{0_\textrm{cs}} + e^{i\phi} \sin\frac{\theta}{2} \ket{1_\textrm{cs}}~,
\end{equation}
$\theta$ and $\phi$ characterises the encoded qubit information.  After dephasing, the logical fidelity of a pure-state cat-code qubit is given by the physical fidelity between the initial and final state, i.e.
\begin{equation}
\mathcal{F}_\textrm{cs}(\theta, \phi)\equiv \bra{\theta,\phi}\rho(t) \ket{\theta,\phi}~.
\end{equation}

On the other hand, if the dephased cat-code qubit is considered as a QSP qubit, the logical state is evaluated by Eq.~(\ref{eq:logical_qubit}).  The logical fidelity is computed by
\begin{equation}
\mathcal{F}_\textrm{QSP}(\theta, \phi) \equiv \Tr{\rho_L(t) \rho_L(0)}~,
\end{equation}
where $\rho_L(t)$ is the QSP logical state encoded by $\rho(t)$; the initial logical state is
\begin{equation}
\rho_L(0) = \frac{1}{2}\left(\hat{I}_L + \cos\phi \sin\theta \hat{X}_L + \sin\phi \sin\theta \hat{Y}_L + \cos\theta \hat{Z}_L \right) .
\end{equation}

\begin{figure}
\begin{center}
\includegraphics[width=\linewidth]{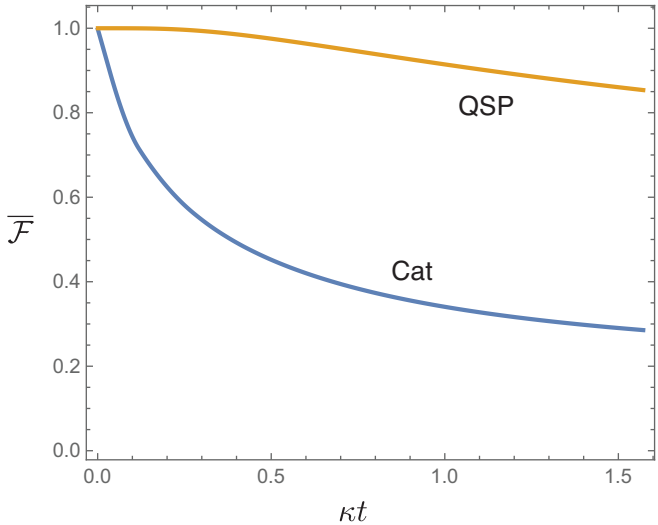}
\caption{ Average fidelity of a cat-code qubit ($\alpha = 2$) after dephasing time $t$.  (Blue)  $\overline{\mathcal{F}}_\textrm{cs}$, the dephased physical state $\rho(t)$ is considered as a pure-state cat-code qubit.  (Orange) $\overline{\mathcal{F}}_\textrm{QSP}$, the same physical state $\rho(t)$ is considered as a QSP qubit.
\label{fig:dp_fidelity}} 
\end{center}
\end{figure}

To compare the dephasing tolerance of the encodings, we consider the average logical fidelity over all qubit state:
\begin{equation}
\overline{\mathcal{F}}_\textrm{code} \equiv \frac{1}{4\pi}\int_0^{2 \pi} \int_0^\pi \mathcal{F}_\textrm{code} (\theta, \phi) \sin\theta d \theta d\phi~,
\end{equation}
where {\small code} $\in\{\textrm{cs, QSP}\}$.  A typical result is shown in Fig.~\ref{fig:dp_fidelity}.  The average fidelity is generally preserved for a longer time if the dephased state is considered as a QSP qubit.  

The intuition behind the improved dephasing tolerance can be analytically understood in the large displacement regime, i.e. $\alpha \gg 1$.  For cat code, a rotation with angle $\varphi$ will displace each coherent state component by a magnitude $\alpha |1- e^{-i \varphi}| \approx \alpha |\varphi|$.  When $ \alpha |\varphi| \gtrsim 1$, the encoded qubit state will be transformed outside the encoding subspace.  Therefore, the physical fidelity can only be preserved for a time $t_\textrm{cs}$, when the angle variance lies within
\begin{equation}\label{eq:t_cs}
 \kappa t_\textrm{cs} =\overline{\varphi^2} \lesssim 1/\alpha^2~.
\end{equation}

On the other hand, if the dephased state is considered as a QSP qubit, we first recognise that the dephasing process does not alter the logical $Z$ information (computational value), i.e. $\Tr{\hat{Z}_E \rho(t)} = \Tr{\hat{Z}_E \rho(0)}$, because rotation commutes with the parity operator, i.e. $[e^{-i \varphi \hat{a}^\dag \hat{a}},\hat{\mathcal{P}}]=0$.  

\begin{figure}
\begin{center}
\includegraphics{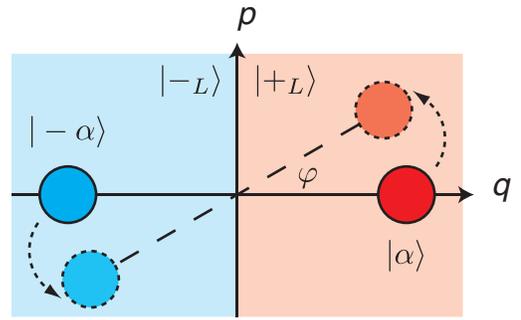}
\caption{ Schematic explanation of dephasing tolerance of logical $X$ information.  Upon random rotation caused by dephasing, the $X$ basis of cat code, $\ket{\pm_\textrm{cs}}\approx \ket{\pm \alpha} $ (solid circles), are displaced outside the computational subspace when $|\varphi| \gtrsim 1/\alpha$ (dashed circles).  
On the other hand, for $|\varphi| \lesssim \pi/2$ the coherent states remain in the same side of phase space (coloured area).  
\label{fig:dp_explain}} 
\end{center}
\end{figure}

The logical $X$ information (coherence) of a QSP qubit is the probability to find the physical state at positive or negative $q$-quadrature (i.e. its Wigner function residing in the left or right side of the phase space).  Although rotation does not commute with the quadrature-sign operator $\hat{X}_E$, a rotated coherent state lies within the same side of the phase space for a wide range of angle $|\varphi| \lesssim \pi/2$.  An illustration is shown in Fig.~\ref{fig:dp_explain}.  
Therefore, the logical $X$ information of QSP encoding is well preserved, i.e. $\Tr{\hat{X}_E \rho(t_\textrm{QSP})} \approx \Tr{\hat{X}_E \rho(0)}$, for a time $t_\textrm{QSP}$ that the angle variance lies within
\begin{equation}\label{eq:t_QSP}
 \kappa t_\textrm{QSP} = \overline{\varphi^2} \lesssim \pi^2/4~.
\end{equation}
When comparing Eqs.~(\ref{eq:t_cs}) and (\ref{eq:t_QSP}), we can see that QSP encoding could preserve logical $X$ information for a longer dephasing time when $\alpha \gtrsim 2/\pi$.

We note that apart from dephasing, QSP encoding can also improve the tolerance of other errors, such as a displacement fluctuation along only one quadrature \cite{Transduction_interference}.  This is the dominating error in some quantum transducer architectures, when measures (e.g. measurement or injected squeezing) can be introduced to suppress one but not both quadrature noise \cite{Higginbotham:2018ca, Transduction_interference}.

\section{Conclusion \label{sec:conclusion}}

In this work, we propose a new mixed-state encoding for quantum computing with bosonic systems.  We first provide a new formalism, which defines a mixed-state encoding in terms of the physical operators that represent logical Pauli operators.  This formalism allows efficient evaluation of logical information encoded in a general physical state.  We then introduce our new quadrature-sign parity encoding, which represents logical computational values by the physical state parity, and logical coherence by the sign of quadrature variable in the physical wave function.  We show that all logical operations required for universal quantum computation, i.e. basis initialisation, logic gates, and information readout, can be implemented by physically feasible processes.  

When comparing with the only known CV mixed-state encoding \cite{kero16}, QSP encoding provides three advantages: QSP logical basis can be deterministically initialised from thermal equilibrium, each logical qubit consists of only one qumode, and the logical measurement can be implemented by CV homodyne detection.  These features enhance the prospect of implementing quantum computers with physical platforms which ground-state cooling is challenging or resources demanding.  Furthermore, even for the systems that cooling is efficient, QSP encoding can improve the noise tolerance of the encoded information.

Generally, CV mixed-state encoding is an area that deserves further exploration.  It is likely that new encodings can be developed to tackle specific implementation limitations, or improve tolerance against structured noise.  Our work provides new techniques and formalism along this direction; the development of hybrid DV-CV quantum computers can thus be facilitated \cite{Andersen:2015dp}.

\section*{Acknowledgement}
 
H.-K. L. thanks University of Toronto and Max Planck Institute for the Physics of Complex Systems for their hospitality, and Dan Sun for her useful comments about refrigerating facilities.  K.M. and D.F.V.J. acknowledge support from NSERC.  A.P. is supported by the Linz Institute of Technology project CHARON.  H.-K.L. acknowledges support by the Croucher Foundation, and by the AFOSR MURI FA9550-15-1-0029. 
 
 \appendix
 
\section{Pauli operators \label{app:Pauli}}
 
For a computational basis pair $\{\ket{\Psi_0}, \ket{\Psi_1} \}$, the pure-state identity and Pauli operators are given by
\begin{subequations}
\begin{eqnarray}
\hat{\mathcal{Q}}_0 \equiv \hat{I}  & \equiv & \ket{\Psi_0} \bra{\Psi_0} + \ket{\Psi_1} \bra{\Psi_1}~, \\
\hat{\mathcal{Q}}_1 \equiv \hat{X} & \equiv & \ket{\Psi_0} \bra{\Psi_1} + \ket{\Psi_0} \bra{\Psi_1}~, \\
\hat{\mathcal{Q}}_2 \equiv \hat{Y} & \equiv & -i \ket{\Psi_0} \bra{\Psi_1} + i \ket{\Psi_0} \bra{\Psi_1}~, \\
\hat{\mathcal{Q}}_3 \equiv \hat{Z} & \equiv & \ket{\Psi_0} \bra{\Psi_0} - \ket{\Psi_1} \bra{\Psi_1}~.
 \end{eqnarray}
\end{subequations}

\section{Equivalence of APO definiton \label{app:equivalence}}

In Sec.~\ref{ssec:representation}, we suggest that APO can be defined by Hermitian operators that obey Pauli algebra (c.f. Eqs.~(\ref{eq:APO_commute}) and (\ref{eq:APO_anticommute})), instead of by the decomposition in Eq.~(\ref{eq:XZ_decompose}).  We here show that both definitions are equivalent.  We first recall that the Pauli algebra can be generated if the Hermitian operators $\hat{X}_E$ and $\hat{Z}_E$ obey Eq.~(\ref{eq:XZ}).  By the self-inverse relation, any eigenstate of $\hat{Z}_E$ have eigenvalues $\pm1$.  A set of orthogonal $+1$ eigenstate can be defined as $\{\ket{\psi_0^{(k)}}\}$.  We can then define the other basis set as
\begin{equation}
\ket{\psi_1^{(k)}} \equiv \hat{X}_E \ket{\psi_0^{(k)}}~.
\end{equation}
To verify that these states are $-1$ eigenstates of $\hat{Z}_E$, we can use the anti-commutation relation, i.e.
\begin{equation}
\hat{Z}_E\hat{X}_E \ket{\psi_0^{(k)}} = -\hat{X}_E \hat{Z}_E \ket{\psi_0^{(k)}}= -\hat{X}_E \ket{\psi_0^{(k)}}~.
\end{equation}
It is also straight-forward to check that the two subspaces $\{ \ket{\psi_0^{(k)}}\}$ and $\{ \ket{\psi_1^{(k)}}\}$ are orthogonal, i.e. $\bra{\psi_i^{(l)}}\psi_j^{(k)}\rangle = \delta_{ij}\delta_{kl}$.  

By sandwiching $\hat{X}_E$ and $\hat{Z}_E$ with the encoding space identity,
\begin{subequations}
\begin{equation}
\hat{I}_E = \sum_k \ket{\psi_0^{(k)}} \bra{\psi_0^{(k)}}+\ket{\psi_1^{(k)}} \bra{\psi_1^{(k)}}~,
\end{equation}
the definitive APO can be expressed in the decomposition form of Eq.~(\ref{eq:XZ_decompose}):
\begin{eqnarray}
\hat{X}_E &=& \sum_k \ket{\psi_0^{(k)}} \bra{\psi_1^{(k)}}+\ket{\psi_0^{(k)}} \bra{\psi_1^{(k)}}~, \\
\hat{Z}_E &=& \sum_k \ket{\psi_0^{(k)}} \bra{\psi_0^{(k)}}-\ket{\psi_1^{(k)}} \bra{\psi_1^{(k)}}~.
\end{eqnarray}
\end{subequations}

\section{Exponential-parity gate realisation \label{app:e-parity}}

All previous proposals of CV exponential-parity gate implementation involve auxiliary qumode or qubit \cite{kero16-2, Lau:2017ky}.  Here we discuss an alternative implementation that is based on the universal CV quantum computation approach \cite{Lloyd:1999vz, Sefi:2011df}, which in principle could generate any Hamiltonian efficiently without using ancilla.  The main challenge is that this approach is efficient only if the target Hamiltonian involves a finite order of quadrature operator.  For exponential-parity gate, however, the effective Hamiltonian is an infinite series of quadrature operator ( c.f. Eq.~(\ref{eq:Parity_qexpansion})).

Our strategy is to truncate the infinite series at a finite order while maintaining the gate accuracy.  We first note that if the exponential-parity gate is applied on a Fock state $\ket{r}$, the infinite operator series in Eq.~(\ref{eq:Parity_qexpansion}) will become an infinite series of complex number.  If the number series can be truncated at a finite order $k \leq K(r)$, but still well approximates the value of the infinite sum, the infinite operator series can then be truncated without significantly reducing the gate accuracy.  Such a truncation is possible for this series, which is a sinusoidal function that has a radius of convergence at infinity.  

For a fixed level of accuracy, $K(r)$ is a monotonic function of the boson number $r$.  
If the physical state of a QSP qubit involves mainly the Fock state components with $r\leq r_{\max}$, then the operator series could be truncated at $k_{\max}\equiv K(r_{\max})$.  
Because each term in Eq.~(\ref{eq:Parity_qexpansion}) is number preserving, the population of Fock states would not change if the Hamiltonian is engineered accurately.  Therefore, $r_{\max}$ and $k_{\max}$ can be determined by the initial state, which in our scheme is a displaced thermal state with thermal excitation $\bar{n}$.  

To estimate $r_{\max}$, we first estimate the magnitude of displacement $\alpha$. According to Eq.~(\ref{eq:logical_basis_fidelity}), if the logical infidelity of a displaced thermal state to the QSP coherence basis is negligible \footnote{Here we assume the threshold of negligibility as $1\%$, but the analysis is applicable to other threshold upon simple changes of numerical parameter.}, the minimum displacement has to be
\begin{equation}
\alpha \gtrsim \sqrt{3}\sqrt{\bar{n}+\frac{1}{2}}~.
\end{equation}

Next, we recall that a thermal state can be considered as a Gaussian ensemble of coherent state, i.e.
\begin{equation}
\rho_\textrm{th} = \int \frac{1}{\pi \bar{n}} e^{-\frac{|\beta|^2}{\bar{n}}} \ket{\beta}\bra{\beta}d^2\beta~.
\end{equation}
In this ensemble, negligible population of coherent state will have displacement beyond
\begin{equation}
|\beta| \gtrsim \sqrt{3 \bar{n}}~.
\end{equation}
Therefore, the state fidelity is not significantly affected if this population is not considered.

Combining these arguments, in a displaced thermal state $\rho = \hat{D}(\alpha)\rho_\textrm{th}\hat{D}^\dag(\alpha)$, the coherent state population would be negligible if the displacement is above
\begin{equation}
|\alpha + \beta| \gtrsim 2\sqrt{3}\sqrt{\bar{n}+\frac{1}{2}} ~.
\end{equation}
In other words, the majority of coherent state in a QSP qubit would have mean boson number at most $\lambda$, where
\begin{equation}\label{eq:lambda_max}
|\alpha + \beta|^2 \lesssim \lambda \equiv 12 (\bar{n}+\frac{1}{2})~.
\end{equation}

We now consider the boson number distribution of the coherent state, $\ket{\sqrt{\lambda}}$, which has more bosons than the majority of coherent state population in a displaced thermal state.  Its boson number population follows Poisson distribution.  In our case of interest, where $\lambda \geq 6$, the total population of Fock states with boson number above some $r>\lambda$ is upper-bounded by  \cite{Klar:2000dn, Poisson_limit}, 
\begin{equation}\label{eq:poisson_bound}
P_{\sqrt{\lambda}}(r) \equiv \sum_{s > r } \left| \bra{s_\textrm{F}}\sqrt{\lambda}\rangle \right|^2 \leq e^{-\lambda} \left(\frac{e \lambda }{r}\right)^r~,
\end{equation}
where the subscript F denotes Fock states.  For other coherent states $|\alpha + \beta|^2 \lesssim \lambda$, the Fock state population above $r$ would be even smaller, i.e.
\begin{equation}\label{eq:prob_compare}
P_{\alpha + \beta}(r) \lesssim P_{\sqrt{\lambda}}(r)~.
\end{equation}

We are now in a position to estimate the maximum boson number $r_{\max}$, above which the the Fock state population is negligible in our encoding displaced thermal state, i.e.
\begin{equation}
\sum_{s > r_{\max} }  \bra{s_\textrm{F}}\rho \ket{s_\textrm{F}} <1 \%~.
\end{equation}
From Eqs.~(\ref{eq:lambda_max}), (\ref{eq:poisson_bound}) and (\ref{eq:prob_compare}), we get a pessimistic bound
\begin{equation}\label{eq:r_max}
r_\textrm{max} \approx 2.5 \lambda = 30 (\bar{n}+\frac{1}{2})~.
\end{equation}
In the construction of exponential-parity gate, the series in Eq.~(\ref{eq:Parity_qexpansion}) can then be truncated at an order $k_{\max}$ that satisfies
\begin{equation}\label{eq:k_max}
\left| \sum_{k=k_\textrm{max}}^\infty \frac{(i \pi r_\textrm{max})^k}{k!} \right| \lesssim 1\%~.
\end{equation}

We note that the purpose of this section is to demonstrate the possibility of truncating Eq.~(\ref{eq:Parity_qexpansion}) at a finite order $k_\textrm{max}$, so that exponential-parity gate could in principle be implemented by the universal CV quantum computation approach \cite{Lloyd:1999vz}.  The truncation order $k_\textrm{max}$ is nonetheless far from optimized.  Obtaining a tighter bound by, e.g. considering the full boson number distribution of a displaced thermal state, or optimising the concatenation sequence, is anticipated but beyond the scope of this work.  

We also note that the above method applies to other physical states that the Fock state population is negligible above some boson number $r_{\max}$.  However, it remains an open question if an exponential-parity gate for \textit{any} state can be engineered accurately by the CV quantum computation approach.  This is unlike the ancilla-assisted approach in Refs.~\cite{kero16-2, kero16, eswapexp} that the implemented gate is accurate for every CV state.

\section{Controlled-phase gate \label{app:cphase}}
 
For self-containedness, we present the logic table for controlled-phase gate $\hat{C}_L$:
 \begin{eqnarray}
 \hat{C}_L |0_L 0_L\rangle = |0_L 0_L\rangle ~~ & ; & ~~ \hat{C}_L |0_L 1_L\rangle = |0_L 1_L\rangle ~;\nonumber \\
  \hat{C}_L |1_L 0_L\rangle = |1_L 0_L\rangle ~~ & ; & ~~ \hat{C}_L |1_L 1_L\rangle = -|1_L 1_L\rangle ~.
 \end{eqnarray}

 \bibliographystyle{apsrev4-1}
\pagestyle{plain}
\bibliography{refs}

\end{document}